\documentclass[10pt,journal,compsoc]{IEEEtran}
\usepackage{graphicx}
\usepackage{multirow}
\usepackage{balance}
\usepackage{diagbox}
\usepackage{comment}
\usepackage{url}
\usepackage{balance}
\usepackage{booktabs}
\usepackage{multirow}
\usepackage{siunitx}
\usepackage{xcolor}

\ifCLASSOPTIONcompsoc
  \usepackage[nocompress]{cite}
\else
  \usepackage{cite}
\fi

\ifCLASSINFOpdf
\else
\fi

\begin{document}
\title{Scalable Light-Weight Integration of FPGA Based Accelerators with Chip Multi-Processors}

\author{Zhe~Lin,~\IEEEmembership{Student~Member,~IEEE,}
        Sharad~Sinha,~\IEEEmembership{Member,~IEEE,}
        Hao~Liang,~\IEEEmembership{Student~Member,~IEEE,}
        Liang~Feng,~\IEEEmembership{Student~Member,~IEEE,}
        and~Wei~Zhang,~\IEEEmembership{Member,~IEEE}

\IEEEcompsocitemizethanks{\IEEEcompsocthanksitem The authors Zhe Lin, Hao Liang, Liang Feng and Wei Zhang are with the Department of Electronic and Computer Engineering, Hong Kong University of Science and Technology (e-mail: zlinaf@connect.ust.hk; hliangac@connect.ust.hk; lfengad@connect.ust.hk; wei.zhang@ust.hk). \protect\\
\IEEEcompsocthanksitem The author Sharad Sinha is with School of Computer Engineering at Nanyang Technological University, Singapore (e-mail: sharad\_sinha@ieee.org).}
\thanks{Manuscript received December 1, 2016; revised June 15, 2017; accepted September 4, 2017.}}


\IEEEtitleabstractindextext{%
\begin{abstract}
Modern multicore systems are migrating from homogeneous systems to heterogeneous systems with accelerator-based computing in order to overcome the barriers of performance and power walls. In this trend, FPGA-based accelerators are becoming increasingly attractive, due to their excellent flexibility and low design cost.  In this paper, we propose the architectural support for efficient interfacing between FPGA-based multi-accelerators and chip-multiprocessors (CMPs) connected through the network-on-chip (NoC). Distributed packet receivers and hierarchical packet senders are designed to maintain scalability and reduce the critical path delay under a heavy task load. A dedicated accelerator chaining mechanism is also proposed to facilitate intra-FPGA data reuse among accelerators to circumvent prohibitive communication overhead between the FPGA and processors. In order to evaluate the proposed architecture, a complete system emulation with programmability support is performed using FPGA prototyping. Experimental results demonstrate that the proposed architecture has high-performance, and is light-weight and scalable in characteristics.
\end{abstract}

}

\maketitle
\IEEEdisplaynontitleabstractindextext
\IEEEpeerreviewmaketitle

\IEEEraisesectionheading{\section{Introduction}\label{sec:introduction}}
\IEEEPARstart{N}{owadays}, the desire for low-power and high-performance design has led to the migration of modern computing systems from homogeneous multicore systems to heterogeneous multicore systems, where hardware accelerators (HWAs) are used to speed up computationally intensive applications~\cite{Chung}. Field programmable gate arrays (FPGAs), which feature great flexibility and high computational capability, are promising candidates to serve as HWAs in heterogeneous systems. Recently, FPGAs have been seen increasingly used in industry to enhance the computation capability of chip-multiprocessors (CMPs). For instance, Altera and Intel provide a research platform, HARP~\cite{harp}, which consists of an Altera Stratix-V FPGA and an Intel Xeon E5 processor. Likewise, Xilinx's Zynq platform~\cite{crockett2014zynq} combines a dual core ARM processor with traditional FPGA fabric to form a programmable system-on-chip (SoC).

With the increasing scale of computer systems, the network-on-chip (NoC) has been used as a high-bandwidth and scalable interconnect architecture for large-scale multicore systems~\cite{qian16,lan16,van16}. It is also promising to integrate an FPGA as a heterogeneous core in an NoC-based multicore system as the next-generation heterogeneous system. Nevertheless, most prior research work has focused on the interfacing of off-chip FPGAs and processors~\cite{hubner2011,papad2012,sander2014,weinhardt2013,jacobsen2015} with a limited number of cores through bus-based communication. Moreover, the rapid increase in the resource capacity and variety of FPGAs over the past few years has made it feasible to implement multiple accelerators on a single FPGA. However, there lacks an interface design which supports (1) run time flexibility when a multitude of processors may request many accelerators, and (2) scalability, as multiple accelerators implemented on a single FPGA cannot be accessed independently by mutually exclusive processors. In light of the above consideration, in this paper, we investigate a high-speed, light-weight and scalable interface architecture that loosely couples the FPGA-based HWAs with NoC-based CMPs and allows flexible invocations of HWAs according to the runtime demands of the processors.
Besides this, our proposed interfacing architecture is different from industrial solutions like HARP~\cite{harp} and CAPI~\cite{capi15}, in that it provides a single-die prototyping for heterogeneous systems, where bus based interfacing (e.g. PCIe in CAPI) is not requisite for all kinds of acceleration.

The main contributions of our work are threefold:
\begin{itemize}
\item We exploit a scalable and light-weight interface for the multiple accelerators in an FPGA which are loosely coupled with CMPs. The key design-specific parameters including the number of task buffers, distributed packet receivers and hierarchical packet senders, are investigated to maintain scalability and maximize the performance of the interface architecture integrated in the FPGA. 
\item We propose a hardware accelerator chaining mechanism that allows HWAs to be serially combined together to collaboratively operate as a monolithic but more complex accelerator during run time. This chaining mechanism exploits intra-FPGA communication and thereby obviates the necessity for excessive data transmission between the FPGA and processors. 
\item A full system including CMPs, the FPGA and the NoC is prototyped and emulated under various workload conditions on an FPGA. A software interface for processors to invoke HWAs is also designed to tackle the programmability challenges. The evaluation results demonstrate the high throughput of our proposed design, compared with both AXI-based and shared FPGA cache solutions.
\end{itemize}

The remainder of this paper is organized as follows. Section~\ref{sec:related} reviews the existing work and discusses their limitations in integrating FPGAs in the context of multicore systems. Section~\ref{sec:system} provides an overview of the whole system. Section~\ref{sec:architecture} describes the proposed architecture in detail, while Section~\ref{sec:progHWA} presents the support for programmability. In Section~\ref{sec:experiment}, the full system evaluation results are presented and analyzed. At last, we conclude the paper and discuss future extension of our work in Section~\ref{sec:conclusion}.

\section{Related Work}
\label{sec:related}
Various communication scenarios between an FPGA and processor cores have been studied in recent years. The work in \cite{hubner2011} proposed a system consisting of an ARM microprocessor and a maximum of four accelerators in an FPGA, with AMBA buses as communication channels. The work in \cite{papad2012} presented a system with a PCI express (PCIe) between processors and an off-chip FPGA, which also achieved reconfiguration when necessary. Similarly, work in~\cite{sander2014} and~\cite{weinhardt2013} realized data transmission between an FPGA and processors using a PCIe and AXI interconnect. These interfacing architectures focused on establishing off-chip communication between the FPGA and processors based on existing bus architectures, which are hard to extend to large-scale on-chip multicore systems. In addition, high platform dependence makes these techniques mostly non-portable across different platforms. Most importantly, they do not investigate the support for sharing various accelerators in an FPGA by multiple processors. In contrast, our proposed on-chip interfacing architecture is optimized under a general situation without platform dependence and in which a number of processors can invoke various FPGA-based accelerators. The authors of RIFFA~\cite{jacobsen2015} proposed a series of works where processors access HWAs. The idea of multiple HWAs accessed by different processors is similar to ours; however, they mainly emphasized providing support for different operating systems to gain access to HWAs, without going deep into hardware performance improvement. To the best of our knowledge, ours is the first work targeting optimizing architectural design for interfacing FPGA-based multi-accelerators with NoC-based multicore systems. Furthermore, our work is complementary to accelerator-rich architectures (i.e., the multicore systems with multiple accelerators) where ASIC blocks or CGRAs are distributed individually in an NoC framework as processing elements~\cite{chen2013,hussain2014}.

\section{Full-System Overview}
\label{sec:system}
\subsection{NoC-based multicore system}
\begin{figure}[ht]
\begin{center}
\includegraphics[width=\linewidth]{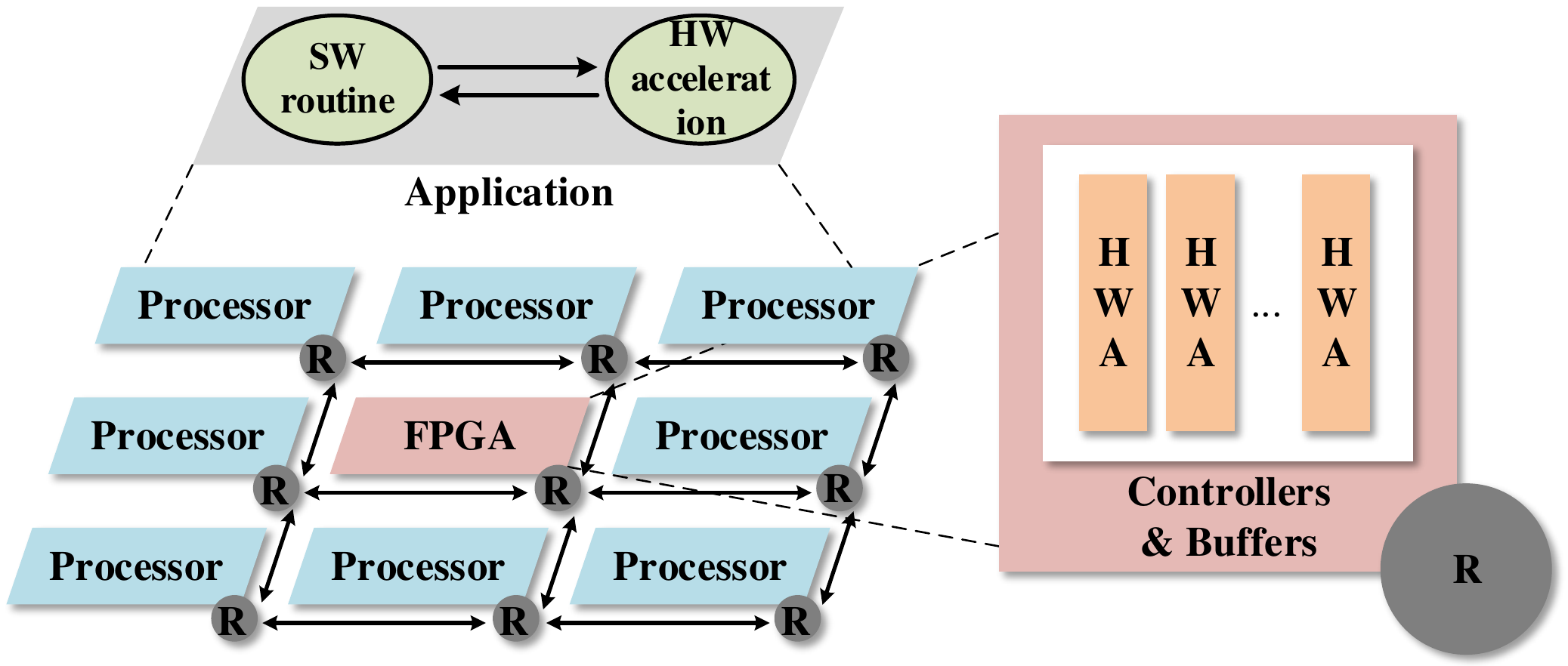}
\caption{Full-system framework.}
\label{fig:mesh}
\end{center}
\end{figure}

NoCs are proposed to be promising on-chip communication architectures for achieving high bandwidth under a limited power budget. The processing elements in an NoC communicate with each other by sending and receiving packets through routers. In the experiments, the employed multiprocessor system-on-chip (MPSoC) architecture is similar to~\cite{girao09}. We adopt a 3-by-3 mesh topology and Fig.~\ref{fig:mesh} presents the system framework. The processors maintain their software routines and leverage the HWAs in the FPGA to complete the acceleration of some computationally intensive works. Note that the difference in sizes between the processors and the FPGA will impact the layout of the chip while it will not influence the topology of the system. In principle, our idea supports any topology with the FPGA placed beside any node. The analysis of the NoC routing algorithms and the traffic patterns~\cite{fu14,wang14} may suggest a specific placement for the FPGA but that is complementary to our main goal and out of the scope of this work.

\subsection{Packet format}
Packet-based transmission is required for an NoC. A packet is composed of several flits: a head flit, multiple body flits and a tail flit, which are the smallest unit in communication~\cite{dally2001route}. We design the flit width to be 137-bit. The head flits are always first to be transmitted in packets and primarily contain routing information together with specific information related to the invoked HWAs. Table~\ref{table:headflit} summarizes the bit information in a head flit. Following the head flits, are the body or tail flits, with bits from 128 to 136 consisting of routing and packet information, and all the remaining bits carry payload data. It is trivial to adjust the flit size for different system configurations by reducing or extending payload bits. Additionally, the number of packets for each HWA invocation is variable since different HWAs require different data sizes, which are distinguished by task head and tail bits.

\begin{table}[ht]
\centering
\small
\caption{Description of the bit index in a head flit.}
\label{table:headflit}
\begin{tabular}{l|l} \toprule
\textbf{Bit index}&\textbf{Description}\\ \midrule
130-136 & Routing information: information for routers\\
128-129 & Packet head \& tail: imply a head/body/tail flit\\
125-127 & Source ID: the requesting processor ID\\
120-124 & HWA ID: the HWA ID to be invoked\\
119 & Packet type: implies a command/payload\\
117-118 & Task head \& tail: imply the first/last packet\\
115-116 & Task buffer ID: implies the task buffer to use\\
113-114 & Chaining depth: HWA chaining times\\
107-112 & Chaining index: 3 HWA chaining indexes\\
105-106 & Packet priority: the arbitration priority\\
103-104 & Packet direction: src/dest of data\\
71-102 & Start address: start address for memory access\\
61-70 & Data size: number of bytes to fetch from mem\\
0-60 & Payload data\\ \bottomrule
\end{tabular}
\end{table}

\begin{figure*}[h]
\begin{center}
\includegraphics[width=17cm]{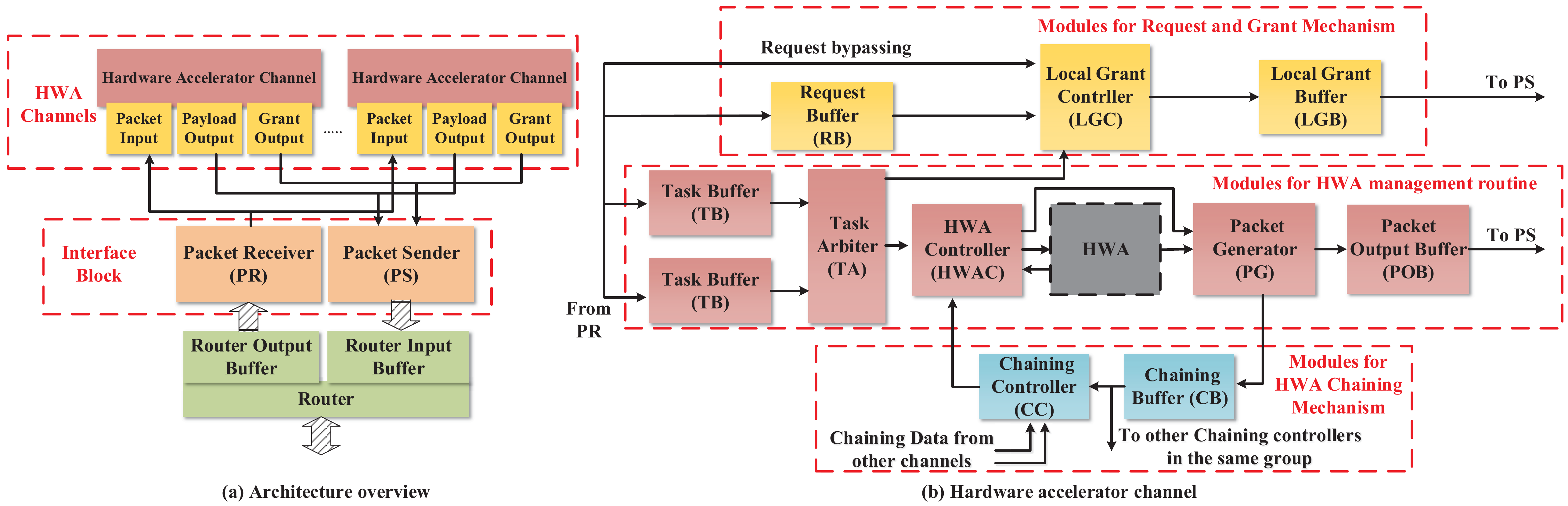}
\caption{(a) Overview of the FPGA-based multi-accelerator architecture; and (b) The detailed design within an HWA channel.}
\label{fig:Overview}
\end{center}
\end{figure*}

\section{FPGA-based Multi-accelerator Architecture}
\label{sec:architecture}
The proposed FPGA-based multi-accelerator architecture is shown in Fig.~\ref{fig:Overview} (a) and comprises interface block and hardware accelerator channels (HWA channels) as the crucial elements. In order to bridge the frequency difference between the FPGA and the local NoC router, router input and output buffers are implemented using asynchronous FIFOs. As there can be multiple HWAs on an FPGA, the scalability of the interface block design is crucial to prevent it from being the performance bottleneck.

\subsection{Interface block}
The interface block is the bridge between HWA channels, the NoC and CMPs. Specifically, it manages the packet transmission and arbitration of both command packets and payload packets. The fundamental components of the interface block include a packet receiver and packet sender which control the packet dispatch and assembly, respectively, between HWAs and the router buffer.

\textit{A.1 Packet receiver (PR)}

The PR reads flits from the router output buffer and dispatches the packets to the corresponding HWA channels. A PR is implemented as a finite state machine which is able to identify different flit types and decode head flit information. It also identifies the packet length in case of a variable-length packet.

\begin{figure*}[ht]
\begin{center}
\includegraphics[width=17cm]{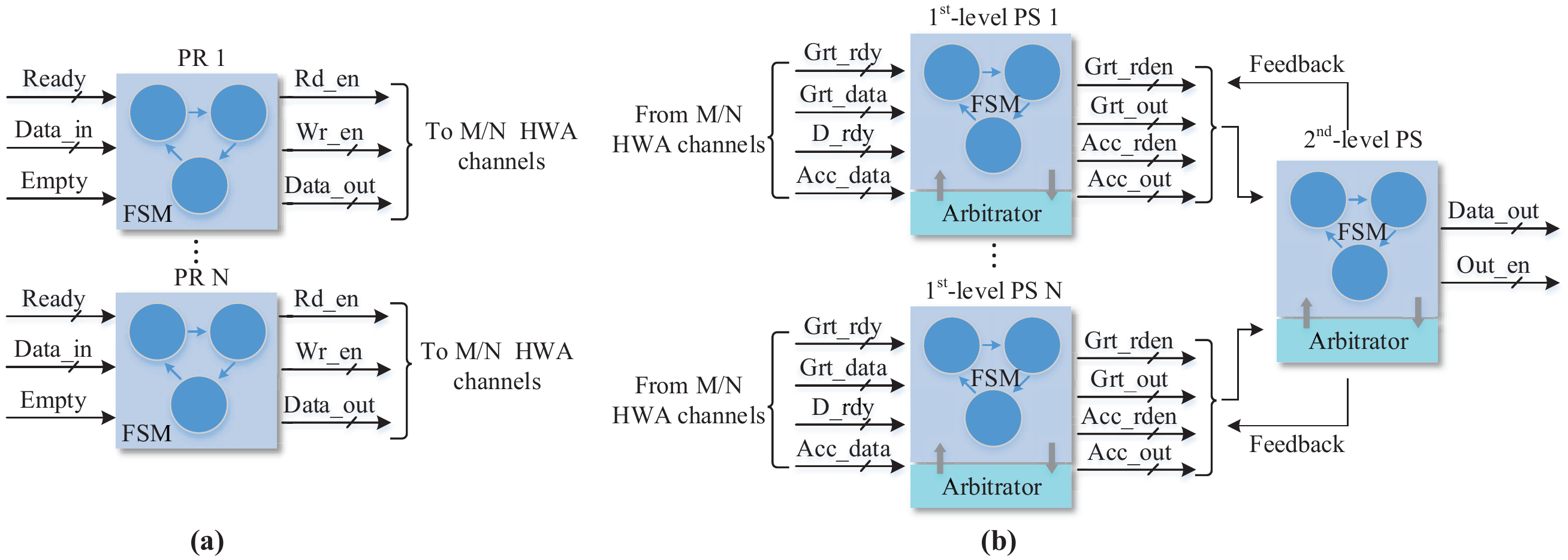}
\caption{(a) Simplified model of distributed PR strategy; and (b) Simplified model of hierarchical PS strategy.}
\label{fig:prcpsc}
\end{center}
\end{figure*}

Considering the case that many HWAs could be implemented and the interface could become the critical path, we explore different design strategies to optimize PR performance: a centralized PR strategy and different distributed PR strategies. For the centralized PR strategy, only a single PR is used to dispatch packets to all the HWA channels, while for the distributed PR strategies, there are multiple PRs, each of which dispatches packets to a fixed number of HWA channels. Fig.~\ref{fig:prcpsc} (a) shows the idea of distributed PR strategies. We investigate various distributed PR strategies and find out the PR strategy with the highest performance by varying the number of PRs. It is observed that the distributed packet receiver strategies can effectively reduce the routing overhead and notably improve the operating frequency of the PR, as demonstrated in Section~\ref{sec:experimentB}.

\textit{A.2 Packet sender (PS)}

The PS arbitrates among different HWA channels and sends the selected output packets to the router input buffer. There are two types of packets to be sent out by the PS: command packets for the HWA requests, and packets for computation results from the HWAs (denoted as result packets). A command packet only has a single flit and it enjoys higher priority in being sent out than result packets. A command packet can be a grant packet which requests input packets, or a notifying packet used to inform the processors of the completion of acceleration. A grant packet would be sent to the requesting processor in the case of a direct access communication scenario or to the memory management unit for a memory access communication scenario, as illustrated in Section~\ref{sec:progHWA}. A round-robin scheme is used to arbitrate command packets from different HWA channels. In contrast, a result packet comprises more than one flit. The PS selects the result packets in a priority-based round-robin manner, with the priority information embedded in the head flits.

By introducing the priority bits in the head flits, the requesting processors can set different priorities for different tasks to be accelerated. This attribute can be removed by setting the priority bits in the head flits as zeros. In such a case, round-robin arbitration is deployed.

Noticing that the complexity of both the arbitration and multiplexing increases with the rise of the number of HWAs, we therefore investigate two types of PS implementation: the global PS strategy and the hierarchical PS strategies. The global PS strategy takes all the command packets and result packets as the input, offers arbitration and sends out the packets. In contrast, the hierarchical PS strategies with the idea shown in Fig.~\ref{fig:prcpsc} (b), clusters a certain number of HWA channels together in the first-level hierarchy and, accordingly, arbitration is done within this specific group. A second-level hierarchical controller finally arbitrates among the first-level hierarchical controllers and then signals the selected first-level hierarchical controller for packet transmission, after which the packet transmission starts.
Experiments are conducted to determine the optimal number of hierarchical PSs to maximize the operating frequency of the FPGA, reported in Section~\ref{sec:experimentB}. The results validate that the optimal hierarchical strategy can significantly reduce the PS delay and as a result, demonstrate a more than 2$\times$ improvement compared with the global PS method.

\subsection{Hardware accelerator channel (HWA channel)}
\textit{B.1 HWA invocation}

Fig.~\ref{fig:Overview} (b) shows the major components necessary to guarantee a robust accelerator invocation. Task buffers (TBs) act as temporary storage for packets with input data for HWAs. Multiple task buffers are desirable to hide the communication delay. The experiment reported in Section~\ref{sec:experimentTB} is conducted to investigate the optimal number of TBs regarding different HWA communication patterns and it reveals that usually two TBs are enough to hide the communication delay. A task arbiter (TA) identifies the ready tasks from the task buffers and selects a task to be executed based on round-robin arbitration. A HWA controller (HWAC) is responsible for reading packets from either task buffers or chaining buffers, and then setting essential control signals to invoke the HWA when the HWA is idle. When the HWA execution is finished, the HWAC will signal the packet generator (PG) to read the execution results. The PG also detects the chaining condition using header information and controls either the packet output buffer (POB) or the chaining buffer to receive the results. If the results are to be sent out, packets are formed simultaneously. The POB serves as temporary storage for result packets before they are granted the chance to be sent back under the supervision of the packet sender.

Note that each HWA's frequency can be different and in order to enable each HWA to run at its own frequency, the task buffers, packet output buffers and chaining buffers are designed to interface between different frequencies. Besides this, the HWAC and PG will work at the same frequency with the HWA to feed the input and generate the output packet under synchronization by asynchronous FIFOs. The control signals crossing different frequencies will be synchronized by two-stage synchronizers implemented by registers.

\textit{B.2 Request and grant mechanism}

Considering the case that a myriad of applications are invoking multiple hardware accelerators in the FPGA, a request and grant mechanism is developed to resolve the contention and ensure the robustness of HWA invocation. For each invocation of an HWA, a request packet would firstly be generated and sent to the FPGA by the processor. The request packet is composed of a single flit with \textit{Packet type ``command''}, \textit{Source ID}, \textit{HWA ID}, \textit{Packet direction}, \textit{Start address} and \textit{Data size}.

As there could be multiple processors requesting the same HWA, a received request packet is firstly queued in the request buffer (RB). A local grant controller (LGC) keeps track of the status of request buffers and task buffers with the support of a status table that is updated every cycle. Based on the task buffers' availability, the LGC generates grant packets in a first-come-first-serve manner, writes the granted task buffer identification into the grant packets and signals the PS for packet transmission. To further reduce the latency for writing and reading requests, a request can bypass the request buffer when no other requests exist in the request buffer. Also note that the grant packets will not be permitted to transmit until a valid task buffer is available.

\textit{B.3 HWA chaining mechanism}

HWA chaining is developed for the case that a task attempts to invoke a series of HWAs sequentially. Notice that the chaining HWAs are pre-specified by the task, and hence the required HWAs can be identified and formed into a chaining group. For instance, in JPEG decompression~\cite{sang2012}, inverse zigzag, inverse quantization, inverse DCT, shift and bound are invoked in a sequence. Therefore, when implemented as HWAs, they can be incorporated in the same chaining group to enable local data reuse among each other, eliminating excessive data transmission through the NoC to the memory or processors. Hence, the HWA chaining mechanism allows a set of HWAs to be invoked collectively in addition to being used individually, making the design more flexible and general.

\textit{Chaining depth} and \textit{Chaining index} are dedicated bits to describe the chaining times and sequences. Moreover, the chaining buffer (CB), chaining controller (CC), HWA controller and packet generator are designed to support chaining. When receiving the results from the HWA, the packet generator first checks \textit{Chaining depth} in the header information. If it is non-zero, chaining is required and both the header information and execution results would be written to the chaining buffer, with the \textit{Chaining length} in the header information decreased by one. The header information in the chaining buffers is transparent to all the chaining controllers in the same group.

The chaining controller is a combinational logic to indicate existing matchings for chaining. It deduces the next chaining \textit{HWA ID} from the \textit{Chaining index} and \textit{Chaining depth} and then compares the derived \textit{HWA ID} to its channel \textit{HWA ID}. It then signals the HWA controller for data fetching when \textit{HWA ID} matchings exist and selects the next chaining buffer to read by a round-robin scheme. The HWA controller then fetches data from the corresponding chaining buffer. The HWA controller prioritizes chaining requests over input requests so as to obviate stalling of an ongoing chained operation and avert overflow of chaining buffers.

Note that other ways to facilitate data reuse on the FPGA are usually through the local cache. Some off-the-shelf commercial designs make use of cache memory to integrate an FPGA and a processor/chip-multiprocessor. Examples of such designs are Intel's Heterogeneous Architecture Research Platform (HARP)~\cite{harp} and IBM's Coherent Accelerator Processor Interface (CAPI)~\cite{capi15}. HARP makes use of a dedicated cache memory implemented on the FPGA. This cache memory is used for shared data communication among the accelerators on the FPGA, and between the FPGA and the applications running on the chip multiprocessor. The CAPI solution is meant for its POWER8-processor-based systems and it allows this system to treat an attached FPGA co-processor as a coherent peer: the FPGA accelerator and the POWER8 system share the same memory space. The POWER8 processor preserves dedicated silicon to implement CAPI. Both HARP and CAPI are implemented using separate boards or sockets for processors and FPGAs. These designs serve well when only one accelerator is implemented on the FPGA. However, for multiple accelerators on the FPGA, there will be heavy memory contention~\cite{choi12}. We also note that Intel's HARP and IBM's CAPI are bus-based designs for different chips, where our design targets FPGA-CMP system-on-chip.

In our architecture, we leverage upon the advantages of block RAMs (BRAMs) in the FPGA to build multiple distributed buffers (i.e., TBs, POBs, etc.) which are used in different stages of HWA invocations. There is no global cache in our design and these distributed buffers abate the potential penalties due to cache misses by buffering data from different stages instead of accessing the cache over and again. When a new set of inputs is demanded, these data are pre-stored by the PR in the TBs and thereby results in a reduced input read-in latency for a HWA compared to the cache access latency. Furthermore, the use of chaining buffers facilitates the communication between grouped HWAs with minimal delay while the communication through the cache tends to take a longer time and cause contention. Generally speaking, our design can support efficient data reuse among chaining HWAs as well as fast input access for separate HWA invocation at the same time. Compared to the AXI-based design as shown in Fig.~\ref{fig:axi} and the shared FPGA cache design as illustrated in Fig.~\ref{fig:cache}, our distributed buffer design reaps the benefits because the input size for each HWA is pre-defined and there is usually little data sharing between HWAs in different chaining groups. Results reported in Section~\ref{sec:expchaining} also demonstrate the overheads in resource and runtime incurred by the chaining mechanism, which are trivial compared to the obtained improvement in performance.

Table~\ref{table:complatency} generalizes the latency for different components in the interface architecture, where $N$ represents the number of flits in the payload packets for a single HWA invocation. The latency incorporates the time for transferring the whole packet with $N$ flits. The buffers (i.e., TB, POB, RB, LGB and CB) are instantiated as FIFOs and therefore they have the same latency for the first payload to be immediately transferred from the input to the output. TA and CC are combinational logics with the delay of a single cycle.

\begin{table}
\begin{center}
\caption{Latency in clock cycles for different components in the interface architecture.}
\label{table:complatency}
  \begin{tabular}[width=\linewidth]{lll}
    \hline
      \multicolumn{2}{|c|}{\textbf{Component}} &
      \multicolumn{1}{l|}{\textbf{Latency (in cycles)}} \\
      \hline
    \multicolumn{1}{|l|}{\multirow{6}{*}{Per HWA}} & \multicolumn{1}{l|}{HWAC} & \multicolumn{1}{c|}{4+N}\\ \cline{2-3}
    \multicolumn{1}{|l|}{} & \multicolumn{1}{l|}{PG} & \multicolumn{1}{c|}{4+N}\\ \cline{2-3}
    \multicolumn{1}{|l|}{} & \multicolumn{1}{l|}{LGC} & \multicolumn{1}{c|}{1}\\ \cline{2-3}
    \multicolumn{1}{|l|}{} & \multicolumn{1}{l|}{TA} & \multicolumn{1}{c|}{1}\\ \cline{2-3}
    \multicolumn{1}{|l|}{} & \multicolumn{1}{l|}{CC}  & \multicolumn{1}{c|}{1}\\ \cline{2-3}
    \multicolumn{1}{|l|}{} & \multicolumn{1}{l|}{Buffers} & \multicolumn{1}{c|}{4+N}\\ \hline
    \multicolumn{1}{|l|}{\multirow{4}{*}{Overall}} & \multicolumn{1}{l|}{\multirow{2}{*}{PR}} & \multicolumn{1}{c|}{Command: 1} \\
    \multicolumn{1}{|l|}{} & \multicolumn{1}{l|}{} & \multicolumn{1}{c|}{Payload: 2+N} \\ \cline{2-3}
    \multicolumn{1}{|l|}{} & \multicolumn{1}{l|}{\multirow{2}{*}{PS}} & \multicolumn{1}{c|}{Command: 1} \\
    \multicolumn{1}{|l|}{} &\multicolumn{1}{l|}{} & \multicolumn{1}{c|}{Payload: 4+N} \\
    \hline
  \end{tabular}
\end{center}
\end{table}
\section{Programmability support for HWA invocation}
\label{sec:progHWA}
\begin{figure}[ht]
\begin{center}
\includegraphics[width=\linewidth]{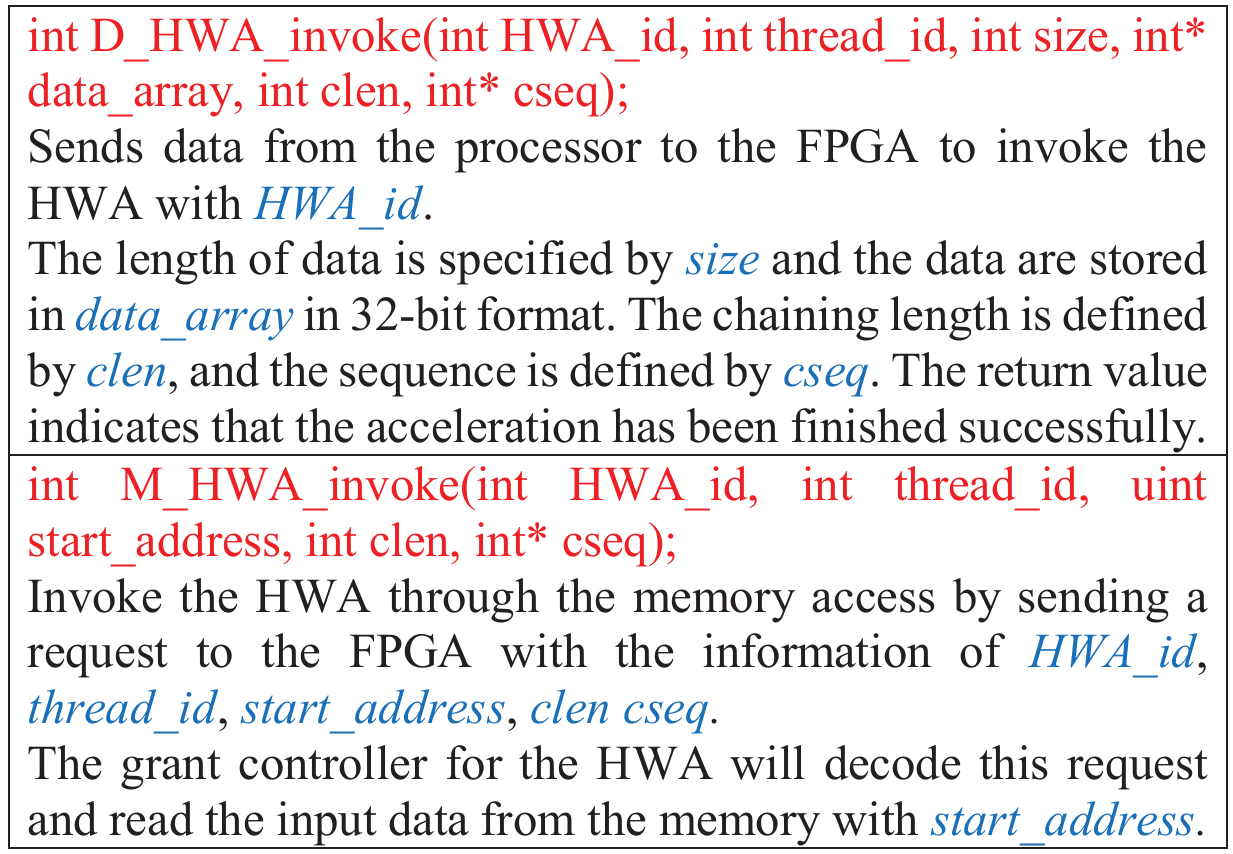}
\caption{Functions for processors to invoke HWAs.}
\label{fig:softcode}
\end{center}
\end{figure}

A software interface is necessary to be developed for processors to invoke HWAs. Specialized C-based functions for HWA invocation are defined and plugged into the user code to specify the information like the \textit{HWA ID} and the caller \textit{thread ID}, as shown in Fig.~\ref{fig:softcode}.

\begin{figure}[ht]
\begin{center}
\includegraphics[width=\linewidth]{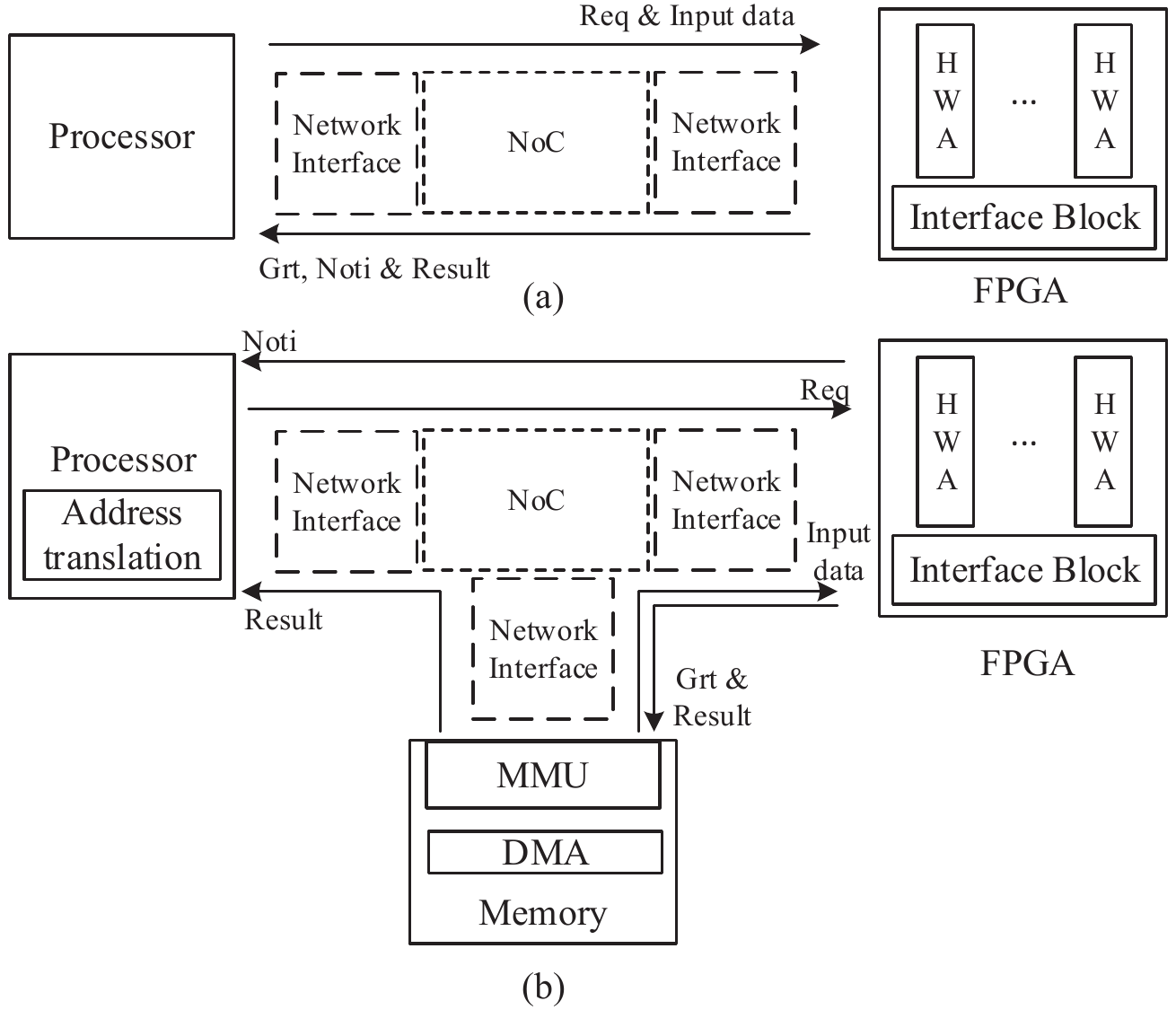}
\caption{Communication scenarios: (a) direct access; and (b) memory access.}
\label{fig:prog}
\end{center}
\end{figure}

More importantly, two communication scenarios between the processors and the FPGA are considered. A processor can either directly send the input data to HWAs or send the requests with the physical addresses of the input data to the HWAs after the virtual to physical translation, as shown in Fig.~\ref{fig:prog}. In Fig.~\ref{fig:prog} (a), the processor directly sends payload packets with input data to the FPGA, while in Fig.~\ref{fig:prog} (b), the HWA can fetch the payload packets from memory through the memory management unit (MMU) by sending the grant packets to the MMU with the specified \textit{Start address} and \textit{Data size} information. When receiving the grant packets from HWAs, the MMU decodes the contained information and initializes data transmission via direct memory access (DMA). In addition, the MMU writes the received result packets in the memory. Notice that the PS is supposed to notify the invoking processor using a packet with the memory address in the header information. Then the processor can fetch data via the MMU either from the memory or the write buffers.

In our design, input packets are received by HWAs and result packets are output and sent over an NoC for processors to process. The data coherency between HWAs and processors is maintained by the processors which invoke the HWAs. Specifically speaking, a processor is responsible for updating the memory and the data coherency state, which are shared among different processors when acceleration results are obtained from the FPGA's HWAs. This is complementary to our proposed design.

\section{Experimental results}
\label{sec:experiment}

\subsection{Experimental setup}
The complete system is prototyped and emulated on a Xilinx Virtex-7 FPGA (xc7vx690tffg1930-3). The NoC is the CONNECT~\cite{papami2012}, with peek flow control, XY-routing and virtual output queues. The employed processor is Microblaze~\cite{xilinx2006mb}, which is commonly used for MPSoC prototyping on an FPGA~\cite{huerta05,singh10,karim14,li14}. We use the Xilinx SDK to compile C-code to execute on Microblaze. Correspondingly, we implement the C-based software interface for accelerator invocation and data communication, as shown in Fig.~\ref{fig:softcode}. Fast simplex links (FSLs)~\cite{rosinger2004} are leveraged for communication between processors and routers. We derive HWAs reported in Table~\ref{table:benchmark} from Xilinx Vivado HLS by performing high-level synthesis of C-based benchmarks from CHStone~\cite{hara2008} and SNU Real-Time Benchmarks~\cite{SNU}, both of which encompass some computationally intensive applications suitable for FPGA implementation. The average lookup table (LUT) utilization is 20424. Three applications use BRAMs and five applications utilize DSPs, showing a variety of resource utilization. We assume that both the CMPs and the NoC operate at 1 GHz according to a commonly used assumption~\cite{park2012}. Note that the 32-bit processor Microblaze implements a classic RISC Harvard architecture exploiting instruction level parallelism (ILP) with a 5-stage pipeline~\cite{barthe}. The pipeline stages (i.e., instruction fetch, instruction decode, execution, memory access and write back) conform to the conventional MIPS pipeline structure. Therefore, Microblaze can be used to extract execution cycles for instructions of classic RISC processors, which are not supposed to change under different operating frequencies. Our proposed multi-accelerator architecture runs at 300 MHz, which approaches the maximum frequency reported in Xilinx Vivado~\cite{feist12}. The HWAs also operate with their maximum frequencies reported by Vivado. Since the whole system is prototyped in FPGA and the FPGA cannot operate at 1 GHz frequency, we scale the frequencies of both the microprocessor and the FPGA according to the ratio expected in a real system to ensure fidelity in emulation, without impacting the key parameters, such as the HWA execution cycle and communication latency, of the system-wide evaluation.

\begin{table}[ht]
\begin{center}
\caption{Benchmark complexity in resources for FPGA implementation.}
\label{table:benchmark}
\begin{tabular}[width=\linewidth]{ccccccc}
    \toprule
    \multicolumn{1}{c|}{\multirow{1}{*}{\textbf{Benchmark}}} & \multicolumn{1}{c|}{\multirow{1}{*}{\textbf{LUT}}} &
		\multicolumn{1}{c|}{\multirow{1}{*}{\textbf{BRAM}}} & \multicolumn{1}{c|}{\multirow{1}{*}{\textbf{DSP}}} & \multicolumn{1}{c}{\multirow{1}{*}{\textbf{FF}}}\\ \midrule
   \multicolumn{1}{l|}{AES Enc} & \multicolumn{1}{c|}{12259} & \multicolumn{1}{c|}{116} & \multicolumn{1}{c|}{0} & \multicolumn{1}{c}{7286}\\
   \multicolumn{1}{l|}{AES Dec} & \multicolumn{1}{c|}{15218} & \multicolumn{1}{c|}{116} & \multicolumn{1}{c|}{0} & \multicolumn{1}{c}{7350}\\
   \multicolumn{1}{l|}{Dfadd} & \multicolumn{1}{c|}{4983} & \multicolumn{1}{c|}{0} & \multicolumn{1}{c|}{0} & \multicolumn{1}{c}{3768}\\
   \multicolumn{1}{l|}{Dfdiv} & \multicolumn{1}{c|}{9661} & \multicolumn{1}{c|}{0} & \multicolumn{1}{c|}{24} & \multicolumn{1}{c}{13171}\\
   \multicolumn{1}{l|}{Dfmul} & \multicolumn{1}{c|}{1927} &  \multicolumn{1}{c|}{0} & \multicolumn{1}{c|}{16} & \multicolumn{1}{c}{2089}\\
   \multicolumn{1}{l|}{Gsm} & \multicolumn{1}{c|}{4257} & \multicolumn{1}{c|}{0} & \multicolumn{1}{c|}{12} & \multicolumn{1}{c}{2643}\\
   \multicolumn{1}{l|}{Prime} & \multicolumn{1}{c|}{161237} & \multicolumn{1}{c|}{0} & \multicolumn{1}{c|}{0} & \multicolumn{1}{c}{277026}\\
   \multicolumn{1}{l|}{Sha} & \multicolumn{1}{c|}{13147} & \multicolumn{1}{c|}{1} & \multicolumn{1}{c|}{0} & \multicolumn{1}{c}{9931}\\
   \multicolumn{1}{l|}{Izigzag} & \multicolumn{1}{c|}{100} & \multicolumn{1}{c|}{0} & \multicolumn{1}{c|}{0} & \multicolumn{1}{c}{98}\\
   \multicolumn{1}{l|}{Iquantize} & \multicolumn{1}{c|}{608} & \multicolumn{1}{c|}{0} & \multicolumn{1}{c|}{76} & \multicolumn{1}{c}{1413}\\
   \multicolumn{1}{l|}{Idct} & \multicolumn{1}{c|}{14552} & \multicolumn{1}{c|}{0} & \multicolumn{1}{c|}{368} & \multicolumn{1}{c}{12390}\\
   \multicolumn{1}{l|}{Shiftbound} & \multicolumn{1}{c|}{7133} & \multicolumn{1}{c|}{0} & \multicolumn{1}{c|}{0} & \multicolumn{1}{c}{7928}\\
   \bottomrule
\end{tabular}
\end{center}
\end{table}

\subsection{Task buffer exploitation}
\label{sec:experimentTB}
Task buffers (TBs) serve as temporary storage between the packet receiver and HWA controller. As a result, increasing the number of task buffers is expected to hide the communication overhead when the HWAs are in operation. In this experiment, we evaluate the optimal number of task buffers to minimize the overall communication latency. Specifically, we evaluate two types of HWAs: (1) an HWA processing a small amount of data with a large execution time (e.g., Dfdiv); and (2) an HWA with an extremely low execution time but working on a relatively large data set (e.g., Izigzag). These two types of benchmarks demonstrate two extreme communication patterns and other HWAs have communication patterns between these two situations.

\begin{figure}[h]
\begin{center}
\scriptsize
\includegraphics[width=\linewidth]{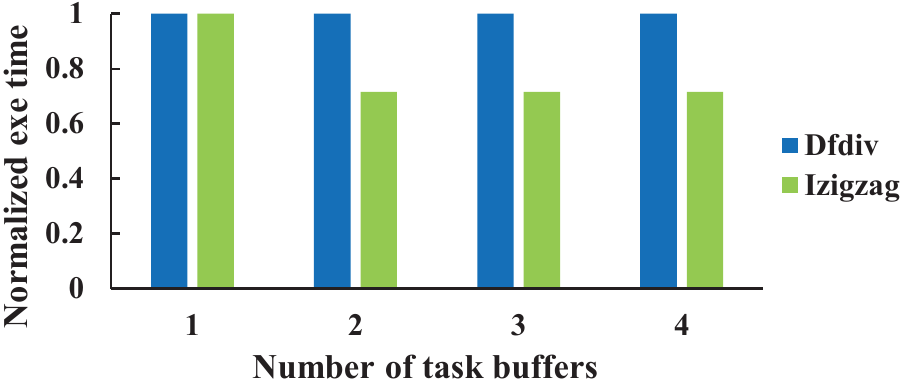}
\caption{The execution time using different numbers of task buffers.}
\label{fig:tb}
\end{center}
\end{figure}

We exploit the total execution time when multiple requests for the same HWA are generated from different processors simultaneously. We record the total
execution time when different numbers of task buffers are utilized to process all the requests. According to the results shown in Fig.~\ref{fig:tb}, there is no
improvement in execution time for Dfdiv as the number of task buffers increases. This is because the time for packet transmission is shorter than the HWA
execution time. In such a situation, new payload packets can be transmitted into the task buffer via the NoC prior to the completion of the last HWA execution and therefore one task buffer is enough. On the contrary, using two task buffers demonstrates a 28.4\% improvement in execution time for Izigzag and no further improvement is observed when increasing the number of task buffers. In such a case, two task buffers are enough to work collaboratively to overlap the packet transmission time with the HWA execution time. These two example HWAs reveal two extremes of communication patterns; hence using two task buffers is sufficient to guarantee high-speed acceleration for various applications. In the following experiments, we incorporate two task buffers for each HWA.

\subsection{Maximum frequency and resource utilization}
\label{sec:experimentB}
\subsubsection{Maximum frequency}
\begin{figure}
\begin{center}
\includegraphics[width=\linewidth]{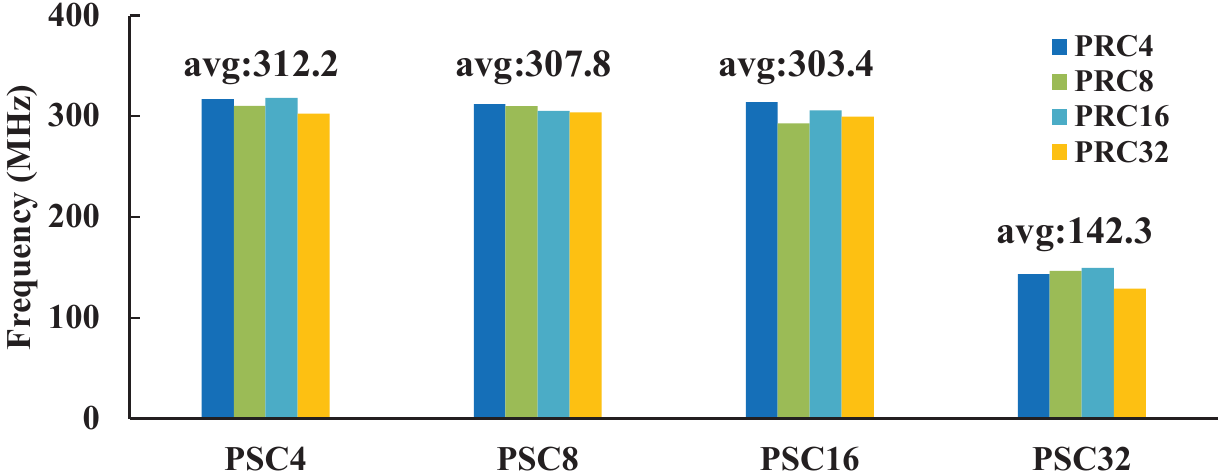}
\caption{Maximum frequency: different PR and PS strategies.}
\label{fig:freq}
\end{center}
\end{figure}

We evaluate the maximum frequencies reported from Vivado 2015.2 after placement and routing with respect to different PR and PS strategies, as shown in Fig.~\ref{fig:freq}. The average maximum frequencies of different PR strategies for the specific PS strategy are shown above the bars. We set the number of HWA channels to be thirty-two, a large number sufficient for scalability investigation. The digits following PR or PS define the number of HWA channels a PR or a first-level PS manages. For example, PS4 indicates that a first-level PS takes control of four HWA channels and correspondingly, the second-level PS arbitrates among eight first-level PSs.

From a PS aspect, the maximum frequencies of all the hierarchical PS strategies are more than twice as high as that of the global PS strategy. Hierarchical strategies remarkably lower the routing efforts because they considerably diminish the fan-in number for both the first-level and second-level designs. Hence, routing congestion resulting from global strategy is alleviated by distributing the heavy centralized routing to multiple paths. Moreover, registers are employed in hierarchical strategies to separate the long wiring in the critical path into two shorter ones. In all, PS4 renders the highest frequency as indicated on top of the bars in Fig.~\ref{fig:freq}, revealing that PS4 best reduces routing congestion and balances delay. Moreover, scalability is preserved under the case of multiple HWAs using PS4.

From a PR aspect, the PR4 strategy surpasses other strategies in frequency, since this strategy trims down the fan-out number of every PR to a desirable value, which similarly lightens the routing burden. PR8 and PR16 provide similar results, while PR32 exhibits the worst performance, since it leads to the heaviest routing burden.

\subsubsection{Resource utilization}
The LUT and BRAM resource breakdown including PRs, PSs and components in HWA channels with dummy HWAs is evaluated using the PR4-PS4 strategy with the highest performance, as shown in Table~\ref{table:resbreak}. The DSP resource is not utilized for any design strategy. Note that TBs and POBs are implemented in BRAMs while other buffers are implemented by distributed memories using LUTs. Furthermore, regarding the different PR and PS strategies we investigated, the LUT utilization ranges between 10.48\% and 10.78\%, and exhibits an average resource consumption of 10.63\% overall and 0.33\% per HWA channel. This value is further verified by implementing a design with eight HWA channels, which utilizes 2.6\% of the resources in all, and again 0.33\% per HWA channel. Therefore, the results validate the light-weight characteristic of our design.

\begin{table}[h]
\begin{center}
\caption{Resource breakdown for the interface architecture in the prototype.}
\label{table:resbreak}
	\begin{tabular}[width=\linewidth]{llllll}
	\toprule
		\multicolumn{2}{c|}{\multirow{2}{*}{\textbf{Component}}} & \multicolumn{2}{c|}{\textbf{LUT}} & \multicolumn{2}{c}{\textbf{BRAM}}\\
		\multicolumn{2}{c|}{} & \multicolumn{1}{c|}{number} & \multicolumn{1}{c|}{\%} & \multicolumn{1}{c|}{number} & \multicolumn{1}{c}{\%}\\ \midrule
		\multicolumn{1}{c|}{\multirow{7}{*}{Per HWA}} & \multicolumn{1}{c|}{TB} & \multicolumn{1}{c|}{100} & \multicolumn{1}{c|}{0.02} & \multicolumn{1}{c|}{4} & \multicolumn{1}{c}{0.27}\\
		\multicolumn{1}{c|}{} & \multicolumn{1}{c|}{TA} & \multicolumn{1}{c|}{2} & \multicolumn{1}{c|}{0} & \multicolumn{1}{c|}{0} & \multicolumn{1}{c}{0}\\
		\multicolumn{1}{c|}{} & \multicolumn{1}{c|}{HWAC+PG} & \multicolumn{1}{c|}{290} & \multicolumn{1}{c|}{0.07} & \multicolumn{1}{c|}{0} & \multicolumn{1}{c}{0}\\
		\multicolumn{1}{c|}{} & \multicolumn{1}{c|}{POB} & \multicolumn{1}{c|}{231} & \multicolumn{1}{c|}{0.05} & \multicolumn{1}{c|}{2} & \multicolumn{1}{c}{0.14}\\
		\multicolumn{1}{c|}{} & \multicolumn{1}{c|}{RB} & \multicolumn{1}{c|}{243} & \multicolumn{1}{c|}{0.06} & \multicolumn{1}{c|}{0} & \multicolumn{1}{c}{0}\\
		\multicolumn{1}{c|}{} & \multicolumn{1}{c|}{LGC} & \multicolumn{1}{c|}{139} & \multicolumn{1}{c|}{0.03} & \multicolumn{1}{c|}{0} & \multicolumn{1}{c}{0}\\
		\multicolumn{1}{c|}{} & \multicolumn{1}{c|}{LGB} & \multicolumn{1}{c|}{247} & \multicolumn{1}{c|}{0.06} & \multicolumn{1}{c|}{0} & \multicolumn{1}{c}{0}\\ \midrule
		\multicolumn{1}{c|}{\multirow{2}{*}{Overall}} & \multicolumn{1}{c|}{PR} & \multicolumn{1}{c|}{870} & \multicolumn{1}{c|}{0.2} & \multicolumn{1}{c|}{0} & \multicolumn{1}{c}{0}\\
        \multicolumn{1}{c|}{} & \multicolumn{1}{c|}{PS} & \multicolumn{1}{c|}{5039} & \multicolumn{1}{c|}{1.16} & \multicolumn{1}{c|}{0} & \multicolumn{1}{c}{0}\\ \bottomrule
	\end{tabular}
\end{center}
\end{table}

\subsection{Throughput of the proposed architecture}
\begin{figure*}[ht]
\begin{center}
\includegraphics[width=5.8cm]{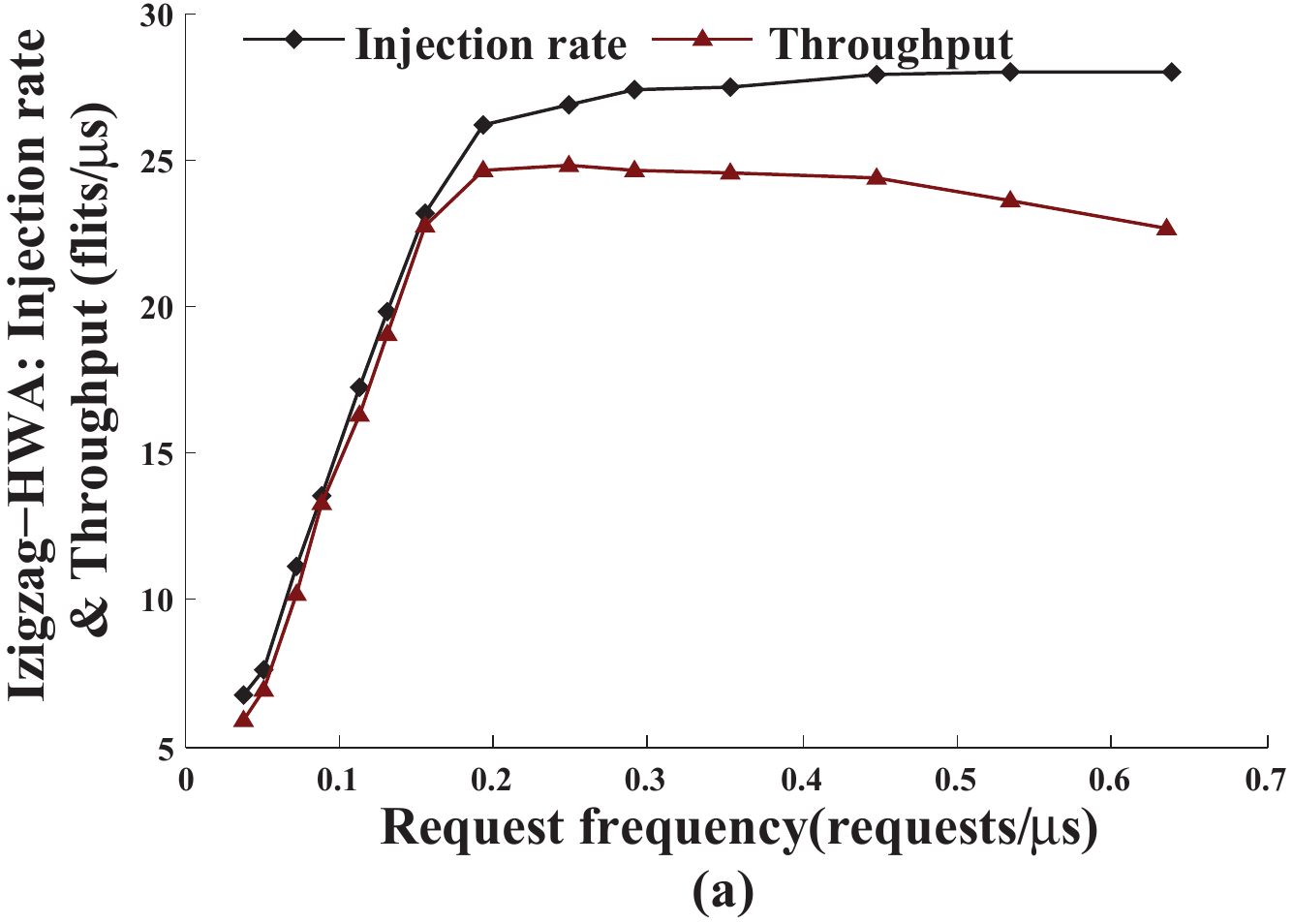}
\includegraphics[width=5.8cm]{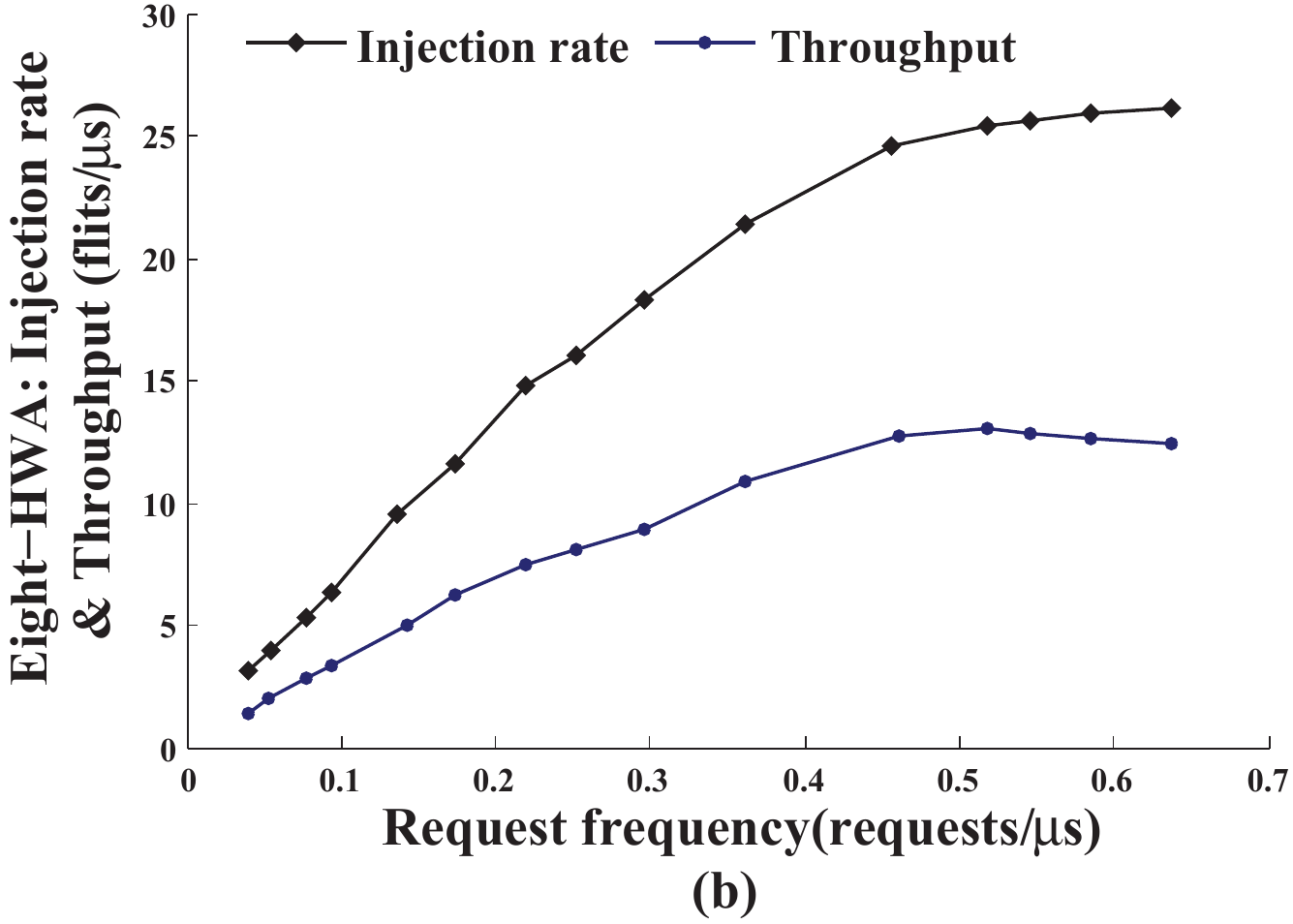}
\includegraphics[width=5.8cm]{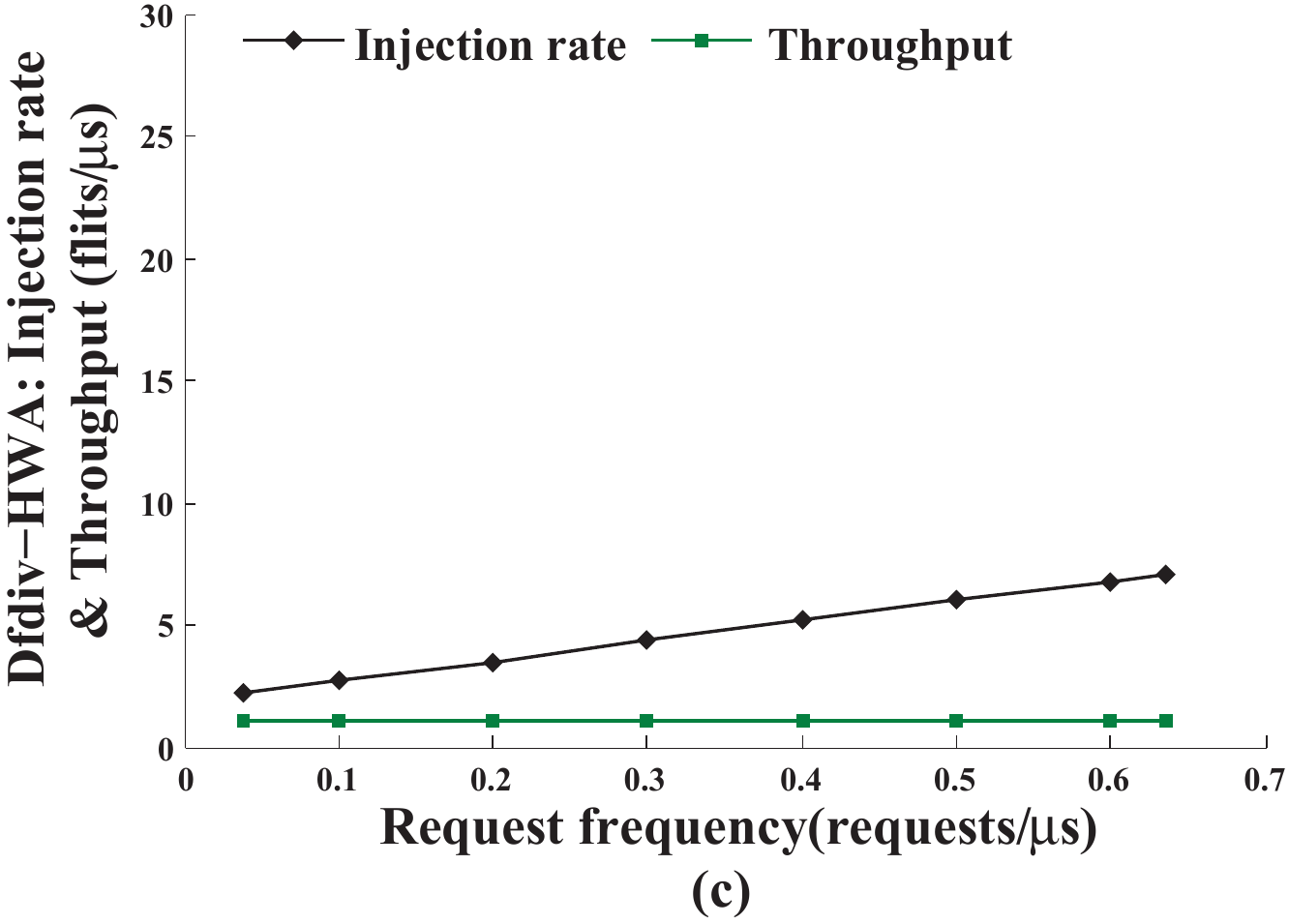}
\caption{(a) Izigzag-HWA injection rate and throughput; (b) Eight-HWA injection rate and throughput; and (c) Dfdiv-HWA injection rate and throughput.}
\label{fig:throughput}
\end{center}
\end{figure*}
To validate the system and evaluate the throughput, we assume eight HWAs on an FPGA and each processor randomly sends requests to specific HWAs under a wide range of request frequencies. Injection rate is used to represent the number of incoming flits per unit time from the router to our design when the full system is at a stable stage. The throughput is calculated as PS output flits per unit time.

In the first case (denoted as Izigzag-HWA), to evaluate the maximum throughput achievable, the eight invoked HWAs are all implemented as Izigzag which has a negligible execution time (i.e., one cycle). Fig.~\ref{fig:throughput} (a) shows both the injection rate and throughput for Izigzag-HWA. The maximum injection rate is 27.95 flits/$\mu$s and the FPGA is busy for 93\% of all the execution time, approaching but not reaching 100 percent owing to the communication overhead incurred by the request and grant mechanism. The throughput becomes saturated at 0.2 requests/$\mu$s and the maximum throughput reaches 24.81 flits/$\mu$s, which is 5.7\% smaller than the injection rate, due to the latency incurred by packet fetching, packet generation and stalling for PS arbitration. When the request frequency further increases, the throughput decreases slightly, because the intensive and substantial data communication eventually accounts for network congestion, which in turn diminishes the data transmission rate.

In the second case (denoted as Eight-HWA), we use the first eight benchmarks in Table~\ref{table:benchmark} with a diversified HWA execution time, to test a common and real scenario. A similar trend to Izigzag-HWA can be seen in Fig.~\ref{fig:throughput} (b). However, the throughput saturates at a higher request frequency, and the throughput is notably lower than the injection rate as a consequence of the non-trivial and diverse HWA execution time.

In the third case (denoted as Dfdiv-HWA), Dfdiv is adopted for all eight HWAs to evaluate the throughput under the other extreme where the HWA execution time is the major dominant factor. As shown in Fig.\ref{fig:throughput} (c), even though the injection rate increases linearly with the rise of request packets, the throughput is chiefly constrained by the lengthy HWA execution time and thereby remains constant.

\subsection{Latency breakdown}
\begin{figure}
\begin{center}
\includegraphics[width=\linewidth]{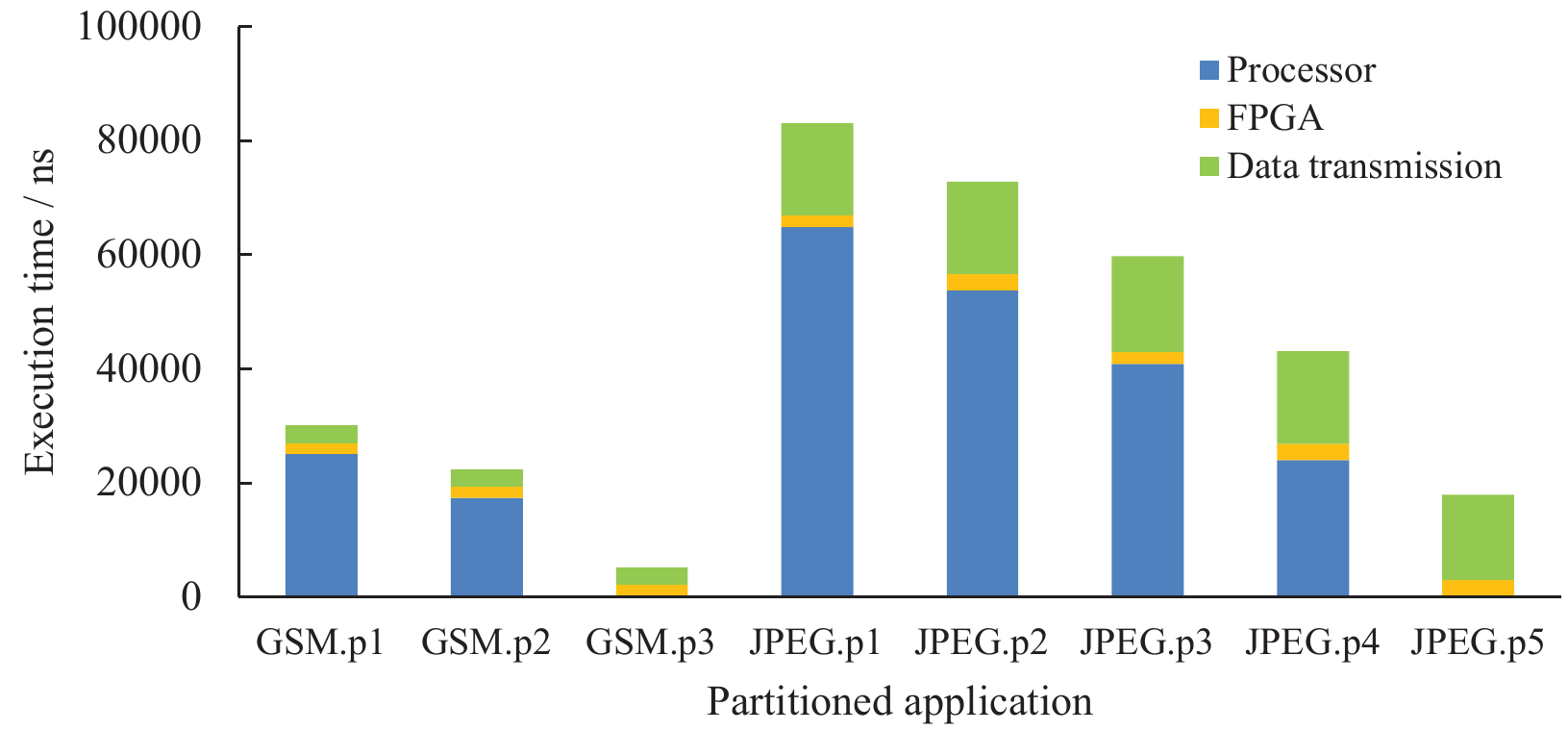}
\caption{Latency breakdowns of different partitions regarding a single invocation.}
\label{fig:pabrk}
\end{center}
\end{figure}

In this experiment, we evaluate the latency breakdown in a single invocation for the processor execution, the FPGA acceleration and data transmission. We conduct task partitioning for the two computationally intensive benchmarks with multiple functions--GSM and JPEG decoder--in Table.~\ref{table:benchmark}. The payload packet sizes are 3-flit for GSM and 18-flit for the JPEG decoder. The latency breakdowns of different partitions are shown in Fig.~\ref{fig:pabrk}. The FPGA executes all the functions in the cases of GSM.p3 and JPEG.p5, which render the smallest overall latency amongst their corresponding partitions. As these two applications incorporate many intensive computations suitable for FPGA acceleration, the improvement in execution time using FPGA acceleration is prominent in all of the different partitions, even with the communication overhead considered. Therefore, the results demonstrate the high efficiency of the FPGA as a platform suitable for accelerating computationally intensive applications in multicore systems.

\subsection{Evaluation of HWA chaining mechanism}
\label{sec:expchaining}
To investigate the efficiency of the HWA chaining mechanism, an experiment is conducted with the Izigzag, Iquantize, Idct, and Shiftbound benchmarks as in Table~\ref{table:benchmark} for JPEG decompression~\cite{sang2012}. These four functions are executed serially to finish decoding for the compressed images in JPEG format. The chaining schemes are: Chaining depth 0 (no chaining), Chaining depth 1 (Izigzag+Iquantize), Chaining depth 2 (Izigzag+Iquantize+Idct) and Chaining depth 3 (Izigzag+Iquantize+Idct+Shiftbound).

\begin{figure}[ht]
\begin{center}
\includegraphics[width=8cm]{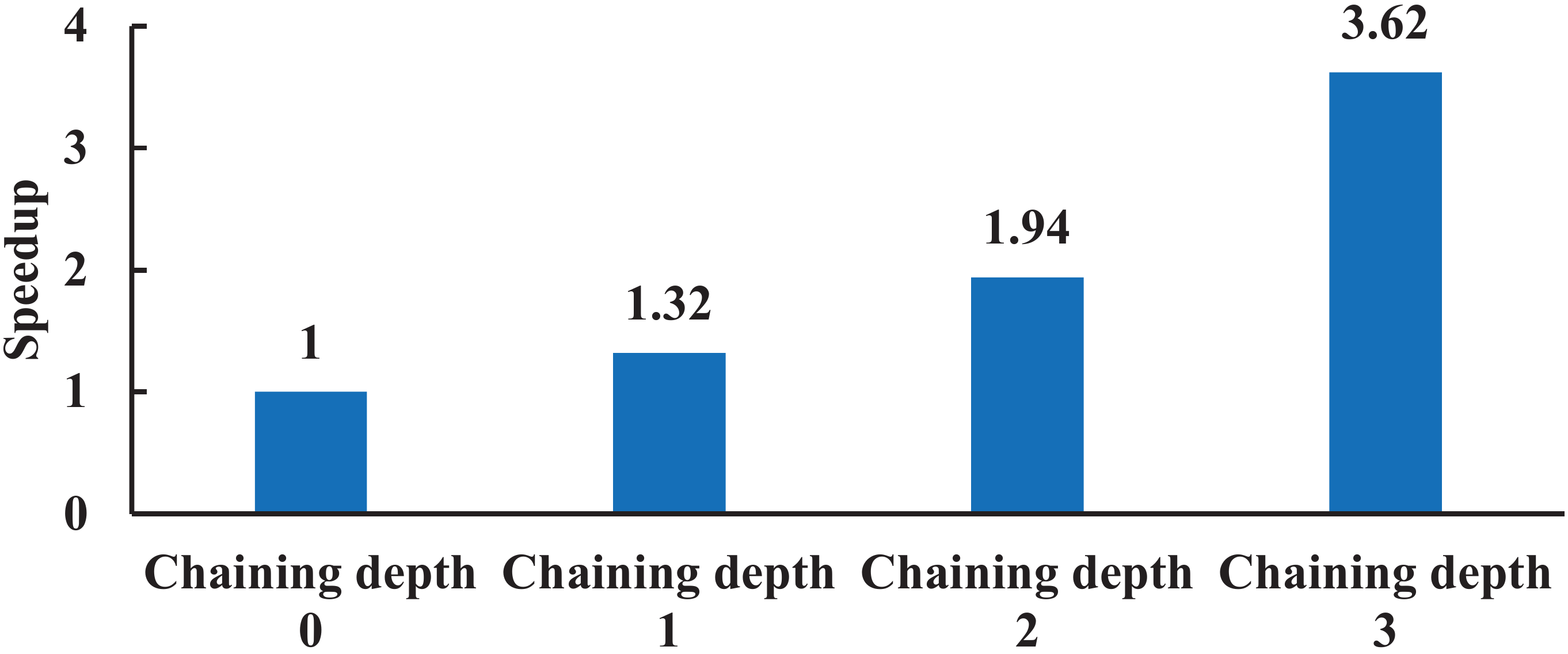}
\caption{Speedup: Different chaining depths v.s. Chaining depth 0.}
\label{fig:chaining}
\end{center}
\end{figure}

The speedup for each different chaining scheme with Chaining depth 0 as the baseline is shown in Fig.~\ref{fig:chaining}. Noticing that the most time-consuming part is the packet sending and receiving operations of the processors, our chaining mechanism effectively diminishes the communication overhead, and it indicates a growing trend for performance improvement when the chaining depth increases.

The communication latency for the chaining mechanism is $4+N$ cycles, where $N$ is the number of result flits to be stored in the chaining buffer. This intra-FPGA communication overhead at runtime is trivial compared with the communication overhead between the FPGA and the processors. Moreover, the LUT overhead per HWA channel for incorporating the chaining mechanism is 526 (0.12\%) and the BRAM overhead is 2 (0.14\%), implying high area-efficiency. As a result, the proposed hardware chaining mechanism demonstrates prominent speedup in the execution time compared with the non-chaining approach, with negligible overheads in runtime and resource, especially when heavy data communication is involved as the chaining depth increases.

\begin{figure}[h]
\begin{center}
\includegraphics[width=\linewidth]{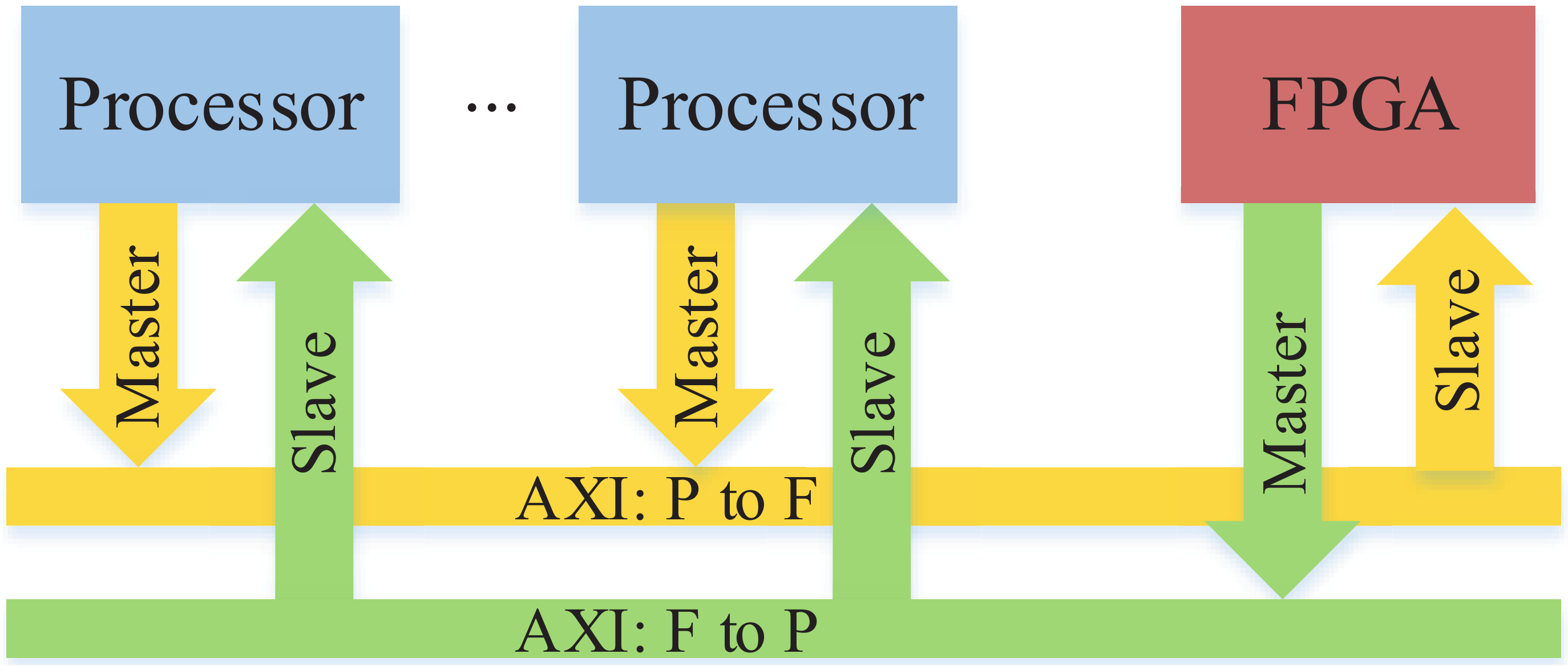}
\caption{System framework based on bus based integration.}
\label{fig:axi}
\end{center}
\end{figure}

\subsection{Comparison with bus based integration}
As illustrated in Section~\ref{sec:related}, the bus-based integration of an FPGA and processors has been studied in prior work. From the industry aspect, the bus-based integration of an FPGA and processors is also extensively deployed in current situation. A representative instance is the ARM Corelink NIC Network Interconnect~\cite{armcl}, which utilizes AMBA AXI4 protocol. In addition, we note that the AXI4 can be well integrated with our proposed interfacing architecture, and in order to perform a fair comparison between bus-based communication and NoC, a prototype is implemented with AMBA AXI4 as a replacement of the NoC in this experiment, as shown in Fig.~\ref{fig:axi}.

The AXI4 frequency is set to be identical to the processors so as to obtain the upper limit of throughput. It is set to be 1 GHz and is scaled to 100 MHz for emulation on the FPGA~\cite{axi04}. The behaviours of injection rate and throughput are similar to NoC based integration. However, as shown in Fig.~\ref{fig:trpcomp}, in comparison with NoC, the maximum throughput for Izigzag-HWA exhibits a reduction of 27\%, while for Eight-HWA, a 53\% decrease in throughput can be observed. For Dfdiv-HWA, the throughput restricted by HWA execution retains an identical constant. Fig.~\ref{fig:prototypecomp} further reveals the communication latency for the AXI-based design and shows a 2.42$\times$ improvement for the proposed NoC design compared with the AXI-based design. In other words, the proposed framework with NoC support indicates predicted advancement in throughput compared with the bus-based integration due to NoC's good scalability, especially when communication overhead becomes the major concern.

\subsection{Comparison with shared FPGA cache}
\begin{figure}[h]
\begin{center}
\includegraphics[width=\linewidth]{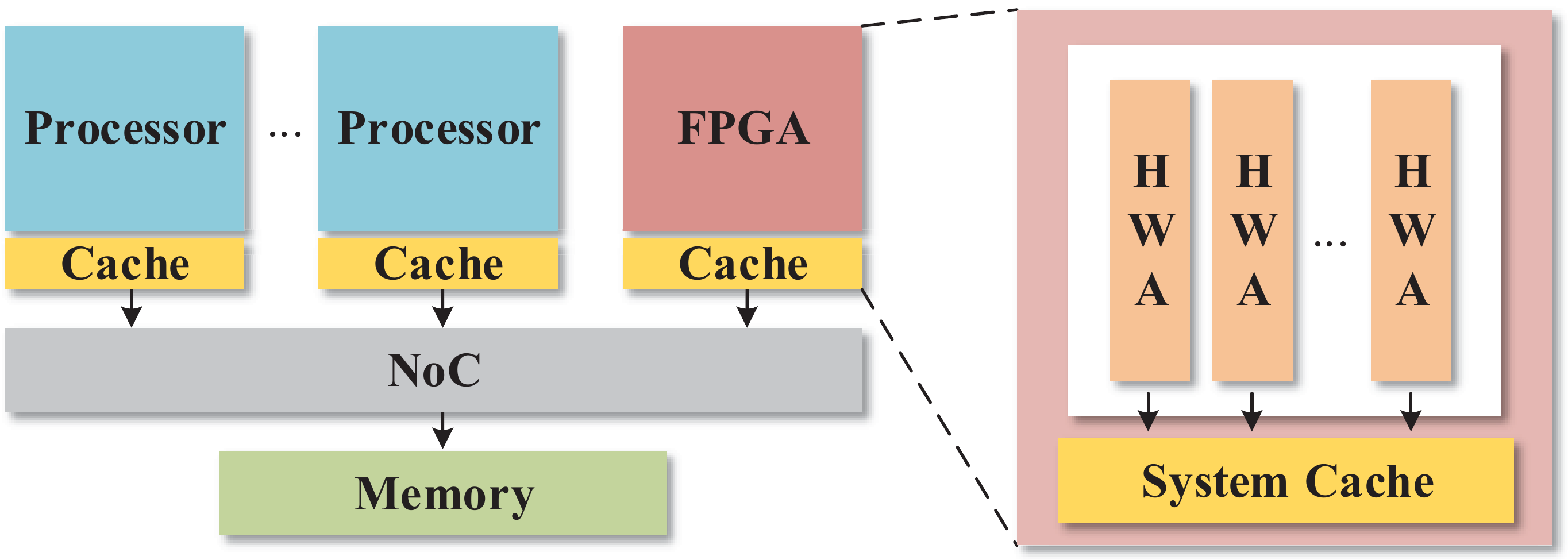}
\caption{System framework with FPGA in-built cache.}
\label{fig:cache}
\end{center}
\end{figure}

\begin{figure}[h]
\begin{center}
\includegraphics[width=\linewidth]{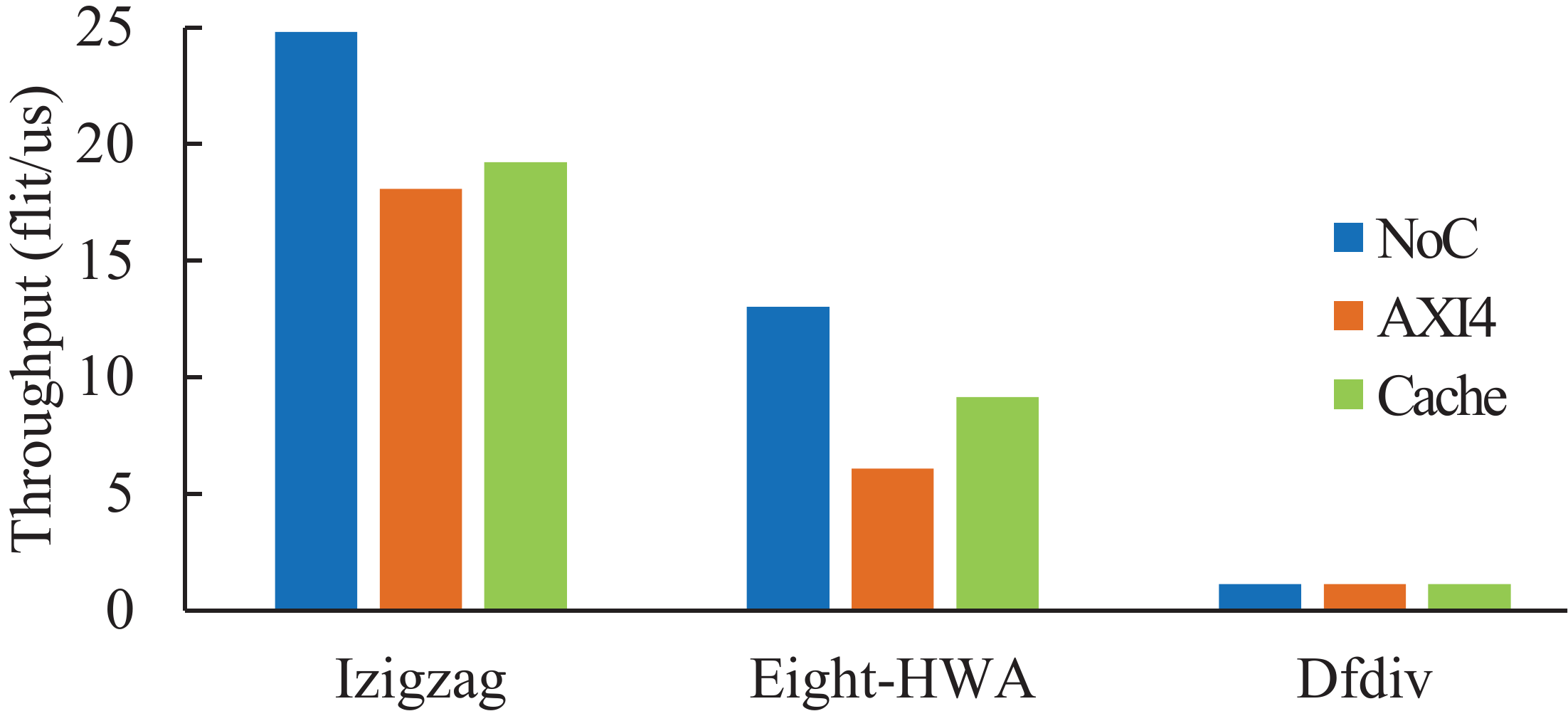}
\caption{Maximum throughput of the three different prototypes.}
\label{fig:trpcomp}
\end{center}
\end{figure}

\begin{figure}[h]
\begin{center}
\includegraphics[width=.9\linewidth]{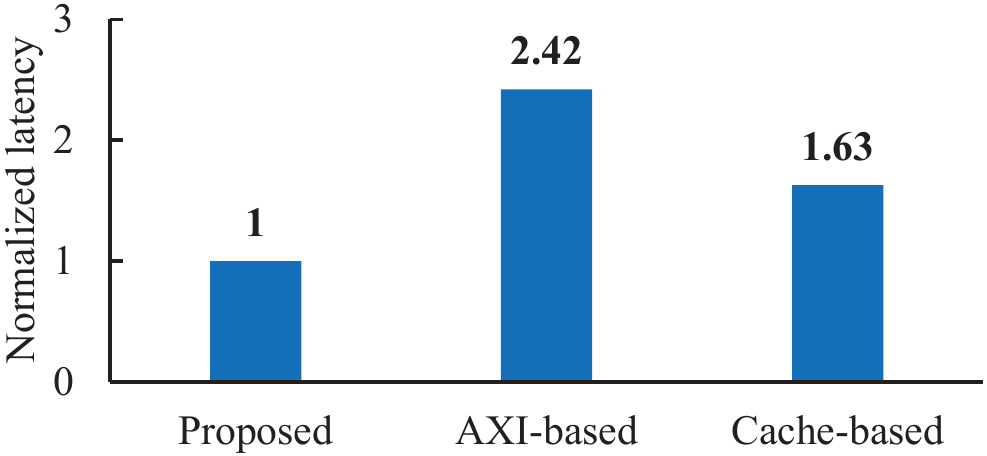}
\caption{Communication latency for a single invocation of the three different prototypes.}
\label{fig:prototypecomp}
\end{center}
\end{figure}

In order to quantitatively characterize the benefits of our design over a shared cache design, we prototype a system using system cache~\cite{syscch} for an FPGA. This cache memory is used to store input and output packets received and sent over the NoC interface, as shown in Fig.~\ref{fig:cache}. The prototyped system is identical to our proposed system but without TBs, POBs and CBs. The HWAs implemented on the FPGA have direct access to the cache. Experimental results shown in Fig.~\ref{fig:trpcomp} indicate a 22.5\% throughput reduction for Izigzag and a 28.2\% reduction for Eight-HWA, compared with our proposed architecture. Fig.~\ref{fig:prototypecomp} also demonstrates an enhancement in communication latency by a factor of 1.63$\times$ for our proposed design in comparison to the shared FPGA cache design. Besides this, the system cache consumes 1\% of the LUT resources and 5\%- 9.5\% of the BRAMs in the FPGA, depending on the cache size, which ranges from 32 Kbyte to 512 Kbyte, with the set associativity fixed as two by default. When there are more chances for HWA chaining, the cache is beneficial. Nevertheless, the intensive access of the cache by all the operating HWAs will cause a surge of congestion and in turn, boost the average access time, which counteracts its merits, thus showing a reduction in throughput compared with our architecture making full use of distributed buffers.
\vspace{4mm}

\section{Conclusion and Future Work}
\label{sec:conclusion}
This paper mainly proposes and implements a platform-independent architectural design for FPGA-based multi-accelerators to efficiently interface with chip-multiprocessors through NoC. Our target is to optimize the performance of the interface when a large number of HWAs are mapped in an FPGA. Specifically, we explore the variations of the key design-specific parameters including: (1) the number of TBs to reduce communication latency; (2) the distributed PR strategies and hierarchical PS strategies to maximize operating frequency as well as maintain good scalability; and (3) the speedup and tradeoff derived from our proposed chaining mechanism. Results show that the optimal set of these parameters can guarantee a more than 2$\times$ improvement in performance. In order to emulate the system-level functionality and evaluate the performance of the proposed interface architecture, we prototype a full system on an FPGA. This prototype encompasses NoC, the FPGA with an integrated interface architecture and multiple HWAs, together with the soft processor cores with HWA invocation functions to tackle programmability issues. We compare our design with commonly used bus-based and FPGA share cache prototypes and finally find our proposed interface architecture demonstrates prominent superiority in performance, area-efficiency and scalability. In our future work, we plan to evaluate the effect of different NoC routing protocols on the performance of the interface.


\ifCLASSOPTIONcompsoc
  \section*{Acknowledgments}
\else
  \section*{Acknowledgment}
\fi
The authors acknowledge the support of the HKUST start-up fund R9336.

\ifCLASSOPTIONcaptionsoff
  \newpage
\fi

\bibliographystyle{IEEEtran}
\bibliography{Ref}

\begin{IEEEbiography}[{\includegraphics[width=1in,height=1.25in,clip,keepaspectratio]{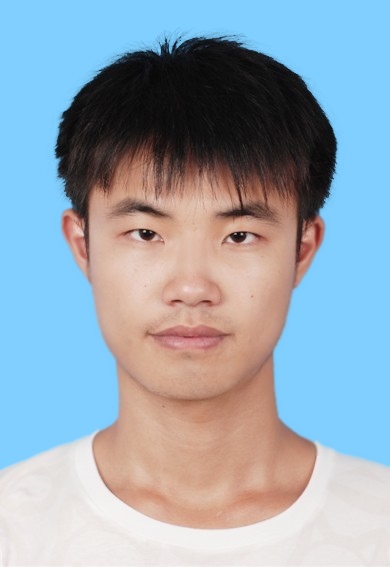}}]{Zhe Lin}
(S'15) received his B.S. degree from School of Electronic Science and Engineering from Southeast University, Nanjing, China, in 2014. Since 2014, he has been a Ph.D. Student in the Department of Electronic and Computer Engineering at Hong Kong University of Science and Technology (HKUST), Hong Kong. Zhe's current research interests cover FPGA-based heterogeneous multicore systems and power management strategies of modern FPGAs.
\end{IEEEbiography}
\vspace{10mm}
\begin{IEEEbiography}[{\includegraphics[width=1in,height=1.25in,clip,keepaspectratio]{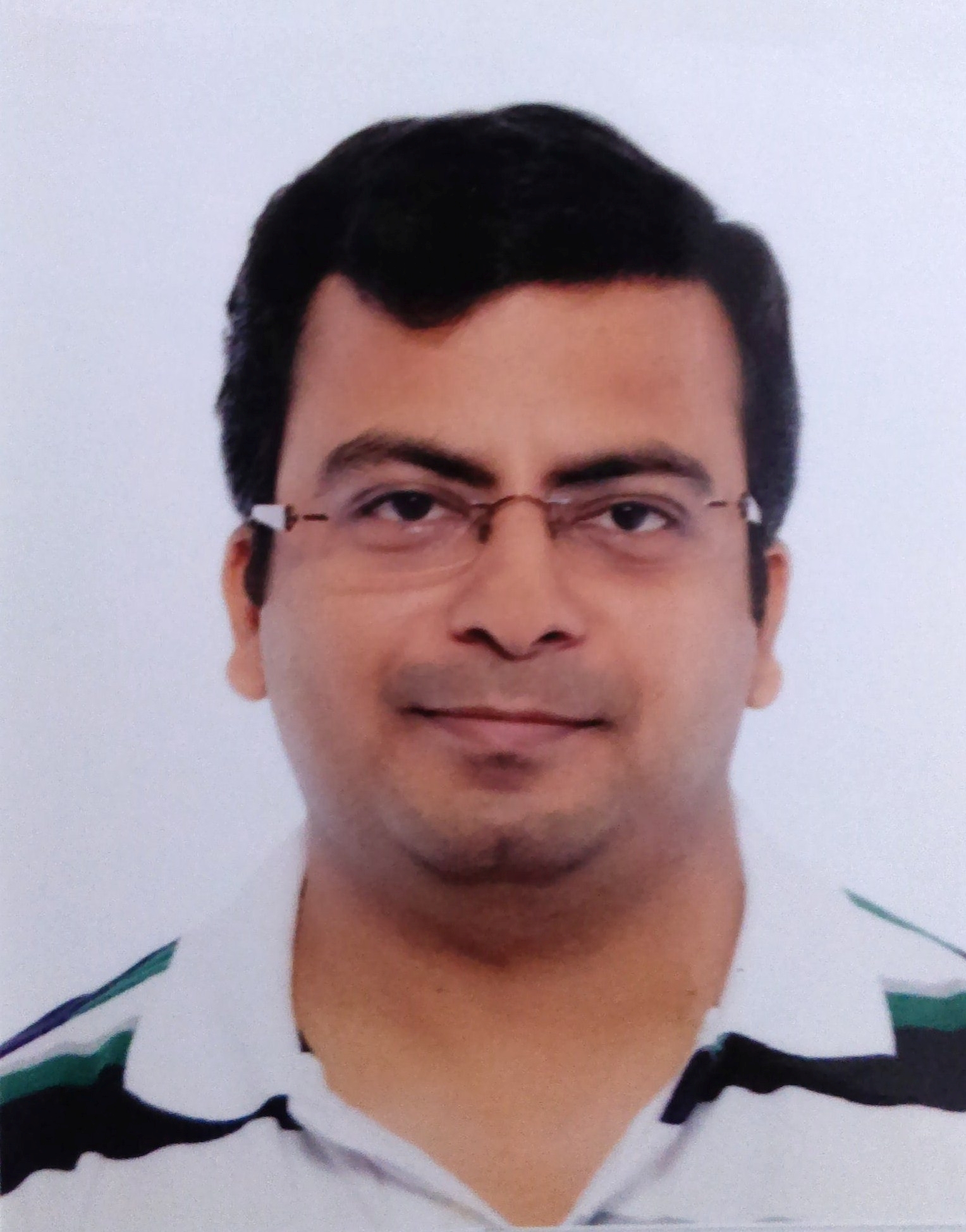}}]{Sharad Sinha}
(S'03, M'15) received his PhD degree in Computer Engineering from NTU, Singapore (2014). He is a Research Scientist in the School of Computer Engineering at NTU. He received the \textit{Best Speaker Award} from \textit{IEEE CASS Society}, Singapore Chapter, in 2013 for his PhD work on High Level Synthesis and serves as a Corresponding Editor for \textit{IEEE Potentials} and an Associate Editor for \textit{ACM Ubiquity}. Dr. Sinha earned a Bachelor of Technology (B.Tech) degree in Electronics and Communication Engineering from Cochin University of Science and Technology (CUSAT), India in 2007. From 2007-2009, he was a design engineer with Processor Systems (India) Pvt. Ltd. Dr. Sinha's research and teaching interests are in computer arhcitecture, embedded systems and reconfigurable computing.
\end{IEEEbiography}

\begin{IEEEbiography}[{\includegraphics[width=1in,height=1.25in,clip,keepaspectratio]{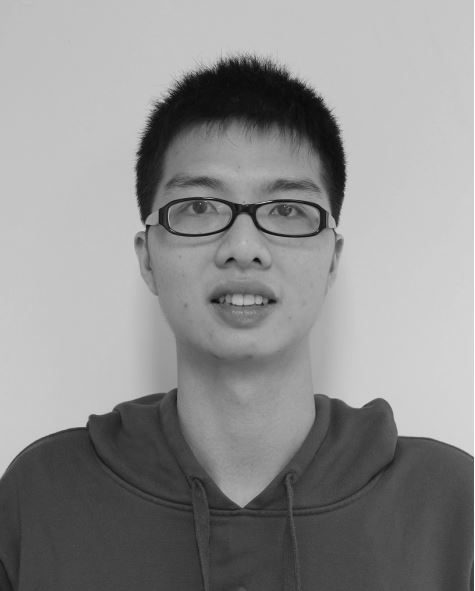}}]{Hao Liang}
received a B.S. degree in software engineering from the Shanghai Jiaotong University, Shanghai, China, in 2011. He is currently pursuing a Ph.D. degree in electronic and computer engineering at Hong Kong University of Science and Technology, Hong Kong. His current research interests include 3-D IC thermal modeling, emerging interconnect technology, embedded systems, and reconfigurable computing.
\end{IEEEbiography}

\begin{IEEEbiography}[{\includegraphics[width=1in,height=1.25in,clip,keepaspectratio]{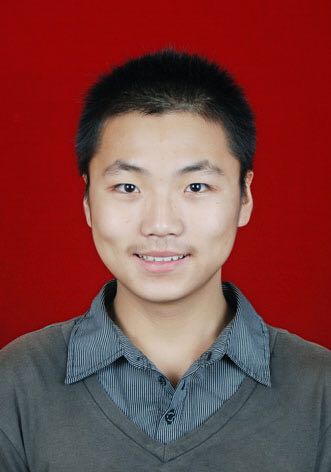}}]{Liang Feng}
received a B.S. degree in Microelectronics from Nanjing University, China, in 2014. He is currently a PhD student in electronic and computer engineering at Hong Kong University of Science and Technology, Hong Kong. Liang's research interests include reconfigurable computing, multi-core system and electronic design automation (EDA).
\end{IEEEbiography}

\begin{IEEEbiography}[{\includegraphics[width=1in,height=1.25in,clip,keepaspectratio]{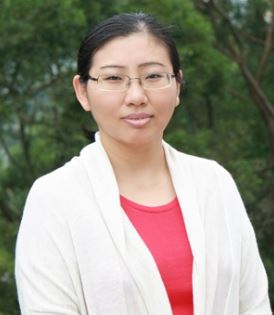}}]{Wei Zhang}
(M'05) received a Ph.D. degree from Princeton University, Princeton, NJ, USA, in 2009. She was an assistant professor with the School of Computer Engineering, Nanyang Technological University, Singapore, from 2010 to 2013. Dr. Zhang joined the Hong Kong University of Science and Technology, Hong Kong, in 2013, where she is currently an associated professor and she established the reconfigurable computing system laboratory (RCSL). Dr. Zhang was a co-investigator of the Singapore-MIT Alliance for Research and Technology Centre, Singapore, where she was involved in low-power electronics. Dr. Zhang was a collaborator with the A*STAR-UIUC Advanced Digital Sciences Center, Singapore, where she was involved in field programmable gate array (FPGA) acceleration for multimedia applications. Dr. Zhang has authored or co-authored over 50 book chapters and papers in peer reviewed journals and international conferences. Dr. Zhang's current research interests include reconfigurable systems, FPGA-based design, low-power high-performance multicore systems, electronic design automation, embedded systems, and emerging technologies.

Dr. Zhang serves as the Area Editor of Reconfigurable Computing of the \textit{ACM Transactions on Embedded Computing Systems} and as an Associate Editor for IEEE TVLSI. She also serves on many organization committees and technical program committees.
\end{IEEEbiography}
\end{document}